# DISCOVERY OF TWO TYPES OF TWINKLING CAN EXPLAIN CONTRADICTORY OBSERVATIONS AMONG TWINKLING ARTIFACT INVESTIGATORS IN ULTRASOUND IMAGING

**Denis Leonov[a,]\*, Nicholas Kulberg[a], Alexandr Gromov[b], Anton Vladzymyrskyy[a], Sergey Morozov[a]**

[a]*Research and Practical Clinical Center of Diagnostics and Telemedicine Technologies, Moscow Healthcare Department, Moscow, 109029 Russia*

[b]*A.I. Yevdokimov Moscow State University of Medicine and Dentistry of the Ministry of Healthcare of the Russian Federation, Moscow, 127206 Russia*

*\*corresponding author; address: 109029, Russia, Moscow, 28-1 Srednyaya Kalintikovskaya str.; phone: +7-909-958-69-38; e-mail: LeonovD.V@yandex.ru*

**Abstract**—Twinkling Artifact is a valuable tool in detecting dense objects such as kidney stones, calculi etc., especially when there is no acoustic shadowing and presence of hyperechogenic tissues obstructs visualization. This phenomenon is not completely understood. Different scientific groups have contradictory findings concerning its properties: some of them observed a decrease in Twinkling intensity at elevated pulse repetition frequencies (PRF), while others found Twinkling to be independent from PRF, etc. In this paper we hypothesize that this kind of contradictions can be partially resolved on an assumption that there are two types of Twinkling. The $1^{st}$ type presumably is produced by random-phased reflections from microcavitation and can be registered even at PRF high enough to suppress blood flow. The $2^{nd}$ is originated from elastic vibrations as the object under investigation swings like a pendulum in the field of ultrasound. These vibrations can be associated with an external source such as a muscles contraction, etc. or the acoustic radiation force from signals emitted by the transducer. The $2^{nd}$ type of Twinkling disappears at high PRF and can be regularly observed in silicone and polyurethane phantoms where the occurrence of cavitation microbubbles is highly unlikely.

INTRODUCTION

Nowadays sensitivity of ultrasound imaging for kidney stones and soft tissue calculi detection is often considered insufficient, which is unfortunate since ultrasound is a method of choice in diagnostics of children, adolescents and pregnant population. In a recent study (Salmaslıoğlu et al. 2018) a range of sensitivity for kidney stones was found to be very broad and dependent on stone sizes with 98% sensitivity for stones larger than 10 mm and 32% sensitivity for stones smaller than 5 mm. Bigger stones typically appear on ultrasound scans as the brightest objects with a distinct posterior acoustic shadow. It is not always the case for smaller stones, which may be indistinguishable from a renal sinus, other fat tissues, tendons and may cast no shadow, thus hindering the diagnostic process. Thankfully, researchers found the way to make diagnostics more reliable, they suggested using ultrasound Doppler Twinkling Artifact, since it was reported to not only facilitate stone diagnostics in complicated situations (Masch et al. 2016; Pabst et al. 2018; Sen et al. 2017; Winkel et al. 2012; Yavuz et al. 2015), but also inform physicians regarding compositions and architectures of stones (Hassani et al. 2012; Jamzad and Setarehdan 2017; Kaya et al.2016; Kim et al. 2010).

Twinkling Artifact (Figure 1) is a well-described phenomenon. It was introduced by Rahmouni et al. in 1996 and defined as a rapidly changing mixture of red and blue behind presumed calcifications in Doppler color flow imaging (CFI). Physicians observed this artifact *in vivo* on kidney stones (Gao et al. 2012; Khan et al. 2017), gallstones (Kim et al. 2010), breast microcalcifications (Fujimoto et al. 2017), epidermoid cysts (Clarke et al. 2016), pleural calcification (Tian and Xu 2018), gastric bezoars (Ahn et al. 2016), testicular microcalcifications (Şekerci et al. 2015), calcified cardiac valves (Tsao et al. 2011), guidewire (Bennet et al. 2015). On the one hand, many researchers agree that Twinkling Artifact is a useful tool for stone detection. On the other, the artifact can obstruct blood flow analysis; therefore, research groups have done some work to suppress the artifact through the signal processing (Leonov et al. 2018a, 2018b; Seo and Wong 2016), others found elevated levels of carbon dioxide capable of eliminating Twinkling (Simon et al. 2017). Apart from that, scientists extensively study the Twinkling phenomenon in phantoms (Haluszkiewicz et al. 2017; Liu et al. 2013a, 2013b; Mavahedet al. 2017; Weinstein et al. 2002) and numerical models (Behnam et al. 2010; Tanabe et al. 2015). However, the results of different studies seem to contradict each other.

## DEPENDENCE ON MACHINE PARAMETERS

Characterization of Twinkling Artifact with various machine settings has been thoroughly investigated by different authors. As shown in Table 1, most of the authors agree that Twinkling tends to become more prominent with both CFI gain and color priority elevated. Carrier frequency and grayscale gain tend to have the opposite effect on the artifact intensity. This finding does not contradict the works of Rahmouni et al. (1996) and Shabana et al. (2009) as they only declared that Twinkling can be observed at all carrier frequencies, the intensity in their studies was not measured. Focal position is recommended to be slightly below the target. It is often noted that Twinkling is less dependent on the focal placement than acoustic shadowing is.

The researchers can be divided in two major groups based on their observation results regarding the pulse repetition frequency (PRF). Some of the researchers (Clarke et al. 2016; Lee et al. 2001; Rahmouni et al. 1996; Shabana et al. 2009) found Twinkling almost indifferent to PRF, thus they recommended to set PRF at its highest to suppress blood signals to avoid potential ambiguity in interpretation when looking for soft tissue calculi and kidney stones. The others (Choi et al. 2014; Naito et al. 2014; Weinstein et al. 2002; Yang et al. 2015) reported the gradual diminishing of artifact with increasing PRF. In one particular paper (Wang et al. 2006) it was observed that Twinkling Artifact was the most intense at the lowest PRF, than its intensity rapidly fell and stayed low with PRF increasing.

Not many studies measured dependence of Twinkling on a wall filter threshold. However, it is a known fact, that this threshold increases with PRF, for this reason we report four papers describing the decline of Twinkling intensity with increase of the wall filtering threshold, the same number of papers found that this threshold had no effect on Twinkling Artifact whatsoever and one study even indicated the rise of intensity with the filtering threshold.

These contradictions made us suppose that there is not just one but, indeed, two types of Twinkling, and the researchers in the field of Twinkling Artifact were unknowingly investigating two different phenomena. These phenomena are easy to confuse as they have a matching appearance on the screen in CFI mode: apparently, stationary objects colored as there is a blood flowing through them (see Figure 1). However, as this study will show, these phenomena have different underlying causes and therefore will be further referenced as the $1^{st}$ and $2^{nd}$ type of Twinkling.

The aim of this study is to test the hypothesis about two types of Twinkling. If two types indeed exist than it might be possible to recognize and differentiate them on the basis of observations with a standard scanner. For deeper understanding, a research machine will be used in our study with the ability to thoroughly observe and analyze raw radio-frequency signals.

## MAIN HYPOTHESES OF TWINKLING

Rahmouni et al. (1996) were the first to report Twinkling Artifact in press. According to their explanation there has to be a slight phase fluctuation between consecutive radiofrequency signals for the Twinkling Artifact to occur. This slight phase fluctuation is further intensified when strong radiofrequency signals are distorted because of the amplifier saturation. However, not all Twinkling objects are the strongest reflectors. Moreover, it is unclear how the slight phase fluctuation occurs in the first place.

Kamaya et al. (2003) in their paper mentioned that conventional Doppler evaluations reflect the velocity of the object based on a frequency shift in the ra-diofrequency signal due to the motion of either the target or the probe. In their experiments, the authors excluded the motion of the probe by securing its position with a ring stand, and, since there was no observable motion of the object under investigation, the machine effects alone were regarded as the cause of Twinkling. They investigated two types of artifact in spectral Doppler mode: the narrowband one that is seen when scanning flat surfaces and the broadband one that occurs on rough exteriors. As a possible reason for the both, the researchers suggested a phase jitter, which can be intensified by rough surfaces. Phase jitter is caused by the slight random time fluctuations in the digital clock that synchronizes the firings. Twinkling Artifact with its wide-band spectrum is generated through slight variations in path length of ultrasound signals.

For Twinkling to occur there has to be not an initial phase fluctuation, but a frequency shift. The way the phase jitter can result in a frequency shift is unclear. It is also unclear how the slight variations in the path can amplify the shift.

Behnam et al. in 2010 investigated the object's motion as the cause of Twinkling. They suggested that the object in the ultrasound field starts oscillating at PRF with the amplitude being several micrometers; an acoustic radiation force, which was investigated in several studies (Andreed et al. 2014, 2016; Fatemi and Greenleaf 1999; Honarbakhsh et al 2010; Nikolaeva et al. 2016), presumably triggered these oscillations. Another team supports the predictions concerning amplitude (Liu et al. 2013a); in their experiments they unsuccessfully tried to observe oscillations with an optical microscope, yet they registered variations of intensity of scattered light with a high-speed camera and a laser when the object was exposed to ultrasound. Thus, they concluded that fluctuations of the object exist and are very small. Yang et al. (2015) also studied the possibility of the acoustic radiation force to cause Twinkling. They measured the force and found it to be insensitive to PRF while Twinkling was highly sensitive and observed at 0.1 and 0.3 kHz and completely disappeared when PRF exceed-ed 0.6 kHz. They concluded that the acoustic radiation force is unlikely to be associated with Twinkling.

Apart from the acoustic radiation force induced by ultrasound machine, object's motion can be caused by internal sources such as heartbeat, breathing, etc., as well as external ones like a vehicle passing nearby, people walking down the hall, etc. One team (Weinstein et al. 2002) even suggested using an additional audio signal to tune in the resonance frequency of the stone to enhance its Twinkling.

The most recent hypothesis (Lu et al. 2013) states that microbubbles in stone crevices could be the dominant cause of Twinkling. The process is known as microcavitation. It supposedly occurs on rough surfaces in diagnostic ultrasound. Microbubbles apparently can be as small as red blood cells.

Polyanskiy and Sapozhnikov in 2018 conducted a numerical simulation to test the 'microbubble hypotheses'. Results of their simulation show that the cavitation of gas bubbles is indeed capable of producing the Twinkling Artifact.

## MATERIALS AND METHODS

**Ultrasound system**

Representation of Twinkling is reported to be machine dependent. That is why in this study we used two standard ultrasound medical diagnostic scanners with a variety of transducers. The first is Sonoace 8000 EX Prime (Medison, Seoul, Korea) with linear probe L5-9EC (Medison, Seoul, Korea), convex probe C3-7ED (Medison, Seoul, Korea) and sector probe P2-5AC (Medison, Seoul, Korea). The second is Sonomed-500 (Spectromed, Moscow, Russia) scanner with linear probe 7,5L37 (Medelcom International, Vilnius, Lithuania), convex probe 3,5C60 (Medelcom International, Vilnius, Lithuania), sector probe 3,5P14 (Medelcom International, Vilnius, Lithuania) and endocavitary probe 6,5CV13 (Medelcom International, Vilnius, Lithuania). We repeated experiments using both ultrasonic machines, observed images in CFI, power Doppler and spectral Doppler modes and compared findings.

In this study, we also analyzed the raw data obtained from the receive path of the research ultrasound scanner Sonomed-500. This machine has an open-architecture which allows the raw data capture from Doppler and B-mode paths for a non-real time processing on a PC and executing third-party program modules to perform real-time experiments on the machine itself. Doppler data went through some preliminary processing stages in the machine including the band-pass filtering and Hilbert transform for creating an analytic signal. The above processing steps were the only "black boxes" in the experiments.

**Data sources**

We used several commercially available ultrasound phantoms: in Gammex 1430 LE Mini-Doppler Flow System (Sun Nuclear Corporation, Middleton, Wisconsin, USA) we could observe conventional Doppler images of a blood mimicking fluid flow as well as mild Twinkling on nylon strings; in ATS 539 (CIRS, Norfolk, Virginia, USA) Twinkling appeared on grayscale targets; in elastography breast phantom by Blue Phantom Twinkling was registered on imitations of breast microcalcifications. In most of the experiments, a clamp held the probe to minimize the effect of random vibrations.

We also investigated Twinkling Artifact with a custom-designed and 3D-printed ultrasound phantom, which is a container with external dimensions of 162x94x82 mm printed with styrene-butadiene-styrene (SBS) (Bestfilament, Moscow, Russia) and having special holdings for targets allowing their placement at predetermined positions. Depending on the experimental requirements, the container was filled with degassed water, ethanol, silicone, or agar-based tissue-mimicking material with the organic sound reflecting content. We degreased all parts with ethanol before pouring water or adding liquid agar. Afterward, we processed them with a hygroscopic surfactant to reduce the residual pocket of gas in the parts' crevices and place a fibrous sound-absorbing material 10 mm thick at the bottom of the phantom to avoid reflections.

We attached the reflecting objects to the custom-built suspensions fixing them in particular positions inside the phantom body. In the early phase of our research, we conducted experiments on a vast number of different materials, but afterward we decided to focus on some minimal set of specimens with known physical properties. The density of the basic material had to be close to that of the bladder stones and calcifications, which is approximately 2.4 g/cm$^3$. We used solid aluminum cylinders with diameter 1.75 mm and length between 8-20 mm ($\rho = 2.7$ g/cm$^3$). Microcalcification was imitated by microcrystals of $CaSO_4$ ($\rho = 2.4$ g/cm$^3$) about 0.1 mm in size, chemically grown in agar jelly.

Some cylinders were scratched with sandpaper to investigate the dependence of Twinkling Artifact and signal characteristics from surface smoothness.

We studied the dependence of Twinkling Artifact on material's density using same-sized cylinders made of iron ($\rho = 7.8$ g/cm$^3$), high-impact polystyrene (HIPS) (Bestfilament, Moscow, Russia) ($\rho = 1.06$ g/cm$^3$), wet wood ($\rho = 0.8$ g/cm$^3$).

Apart from the phantoms, Twinkling was studied in three adult volunteers (a local institutional review board approved the experimental protocol; all the human subjects gave their informed consent for the study):

– Volunteer 1 in the neck area contains 3-mm calculus occurred in the place where hematoma had been previously diagnosed;
– Volunteer 2 has a 15-mm stone located in the right kidney producing acoustic shadowing clearly visible in B-mode and creating pronounced Twinkling Artifact;
– Volunteer 3 has the gallbladder stones producing shadowing and almost no Twinkling.

**Signal model**

In the process of CFI frame acquisition multiple pulses are being send to each point of the region of interest (ROI). The number $K$ of pulses ranges between 8-32 depending on the frame rate needed. Each received pulse represents the state of the system in the current moment of time. In fact, a series of $K$ images is being constructed; these images reflect the changing state of the ROI. For that, a sequence of $K$ ultrasonic pulses with frequency $\omega_0$ and interval $T_{PRF} = \frac{2\pi}{\omega_{PRF}}$ is sent along each beam. In the receiving path, these signals are amplified, digitized and subjected to Hilbert transform to create an analytic signal. Therefore, Doppler data can be represented as a four-dimensional array of complex values $V_{klmn}$, where $k,l,m,n$ are the numbers of the pulse in Doppler sequence, beam, sample and frame, respectively. The set consisting of $K$ complex amplitudes in each point of the ROI represents the ensemble $\mathbf{x} = [x_0, x_1, \ldots, x_{K-1}]^T$. Each element of the ensemble is a complex value, which can be described in accordance with a traditional Doppler signal model:

$$x_k = P_k e^{i\varphi_0} \left( A_k e^{i\omega_A t_k} + B_k \right) e^{i\omega_B t_k} + E_k, \qquad (1)$$

where $P_k$ describes the influence of the scanning impulse envelope at the moment $t_k$, i.e., for a single reflector initially positioned at the center of the envelop $P_k$ is $P\left(\frac{V_A t_k + V_B t_k}{c}\right)$. The scanning impulse can be assumed having Gaussian shape with the duration of several oscillation periods at frequency $\omega_0$; $\varphi_0$ is the initial phase of the ensemble.

$A_k$ is the complex amplitude of blood reflection, $\omega_A = \frac{2V_A}{c}\omega_0$ is a Doppler frequency shift caused by a blood movement with velocity $V_A$, $c$ is a speed of sound, $t_k = kT_{PRF}$ is the interval between zeroth and $k$-th pulse in a sequence, $k = 0, 1, ..., K-1$;

$B_k$ is the complex amplitude of slow-moving tissue signals, usually it is 20–30 dB greater than the blood echoes. $\omega_B = \frac{2V_B}{c}\omega_0$ is a Doppler frequency shift caused by tissue motion and transducer trembling; its velocity is assumed to be $V_B \ll V_A$;

$E_k$ is the complex component representing thermal and quantization noise in receiving front-end modules, its amplitude is 10–15 dB lesser than blood echoes.

In this study we hypothesize that Twinkling occurs because machine receives some unordinary signals from the Twinkling area, which cannot be correctly described within the limitation of the traditional Doppler signal model (1), however, these Twinkling signals possess certain features of blood reflections, and the machine interpret them as blood signals. We suggest that Twinkling signals have unique features and envision the machine could be able to distinguish them from blood if these features were described in machine's language, i.e., preprogrammed in the algorithms. On this assumption, we compliment the traditional Doppler signal model (1) with two components – $C_k$ and $D_k$ – responsible for Twinkling:

$$x_k = P_k e^{i\varphi_0}\left(A_k e^{i\omega_A t_k} + B_k + C_k e^{i\varphi_k} + D_k\right)e^{i\omega_B t_k} + E_k, \qquad (2)$$

where $C_k$ is the amplitude of vibration-related signals leading to the appearance of the 2nd type of Twinkling, its power is 0–10 dB greater than that of the soft tissue signals. Stones can move along with tissues resulting in the Doppler shift $\omega_B$; the acoustic radiation force can trigger their oscillations as well. We consider these oscillations as an important cause of Twinkling Artifact. They are the reason for an additional phase shift $\varphi_k$ to occur. This phase change can be estimated via this equation:

$$\varphi_k = \frac{\omega_0 R}{c} \sin(\omega_C t_k), \qquad (3)$$

where $R \ll \lambda = \frac{2\pi c}{\omega_0}$ is the amplitude of forced oscillations of stone along the scanning beam, $\omega_C$ is the frequency of these oscillations;

$D_k$ is the complex amplitude increment from cavitation resulting in the 1st type of Twinkling. The amplitude of this signal is usually greater than the vibration-related component $C_k$. It is assumed that $D_k$ changes from pulse to pulse chaotically.

**Signal processing algorithms**

The ultrasound machine Sonomed-500 has advanced research capabilities allowing to obtain the raw radio signals from its receive path for further investigations. For processing of these signals, we prepared program modules, including conventional CFI signal processing algorithms. There is a broad range of references for such processing methods, e.g., (Bjærum and Torp 2000; Gerbands 1981; Kargel et al. 2002; Kargel et al. 2003; Leonov et al. 2019a, 2019b; Lo et al. 2008; Løvstakken 2007; Shen et al. 2013; Torp 1997; Wang et al. 2006; Yoo et al. 2003; Yu and Cobbold 2008; Yu et al. 2007; Yu and Løvstakken 2010). We wrote all our program modules in C++ and can either upload them to the machine to analyze data in real time or use them for detailed analysis on a personal computer.

Digital signal processing includes:
– eliminating signals from slow-moving reflectors (wall filter, clutter filter). We tested several different variants of this procedure: bandpass filtering; polynomial regression of different orders; principal component analysis; Karhunen–Loève decomposition; empirical mode decomposition (Leonov et al. 2019a and 2019b). In our experiments we filtered Doppler sequences and found the polynomial regression of 2nd and 3rd orders to be the most suitable for the task;

– smoothing and signal accumulation for noise reduction. Recursive low-pass filters were implemented in two spatial directions along with the frame averaging;

– phase analysis of signals within a Doppler sequence (this information is used in CFI);

– Doppler signals intensity analysis (this mode is known as power Doppler).

We tested the algorithms under the following conditions:

– observation of the liquid flowing along the Doppler phantom's tube; flow velocity was being changed between 0-174 cm/sec. The probe was held in the clamp;

– the same observation, but the probe was held in hand. Wall filter efficiency was evaluated;

– observation of targets, in which Twinkling Artifact appears in standard ultrasound machines. The machine parameters were adjusted for the most evident artifact appearance.

**Experimental design**

*Experiment I: visual representation of Twinkling Artifact.* With two standard medical ultrasound machines in all available Doppler modes, which were CFI, power and spectral Doppler, we investigated the presence of Twinkling in each target in the phantoms at our disposal and the dependence of Twinkling on machine settings: PRF, wall filter type, color write priority, Doppler power, carrier frequency, grayscale gain and focal position to reproduce the findings of other researchers given previously in the 'Dependence on machine parameters' section. The main question of this experiment was: "Can two types of twinkling be reliably differentiated through the observation of their appearance on a screen of a standard scanner?"

*Experiment II: looking for the $1^{st}$ and $2^{nd}$ type of Twinkling signals predicted by mathematical model* (2). Now, after we can reliably detect and differentiate $1^{st}$ and $2^{nd}$ type of Twinkling by their appearance on the screen, we want to look deeper and study the nature of the artifact by looking at raw signals themselves. The model (2) predicted that the signal diagram of the 1$^{st}$ type of Twinkling would be chaotic and the diagram of the 2$^{nd}$ type on a complex plane would look linear – these are the peculiarities we are looking for in this experiment.

*Experiment III: quantitative differentiation of 1$^{st}$ and 2$^{nd}$ type of Twinkling signals.* Differentiating signals based on the visual observation of their diagrams is a tedious process. We introduced a mathematical rule to speed up the process of differentiation: according to the model (2), the 2$^{nd}$ type of Twinkling has a signal of which real and imaginary parts change either in phase or with a 180 degree delay, therefore a rule based on correlation between them can be used for recognizing these signals: if the correlation is close to one then the 2$^{nd}$ type of Twinkling is considered to be detected, else if the correlation is low enough then it is the 1$^{st}$ type, otherwise there could be a mixed type; the correlation is given by:

$$r_{\text{re,im}} = \left| \frac{\sum_{k=0}^{K-1} \operatorname{re} \tilde{x}_k \cdot \operatorname{im} \tilde{x}_k}{\sqrt{\sum_{k=0}^{K-1} (\operatorname{re} \tilde{x}_k)^2 \sum_{k=0}^{K-1} (\operatorname{im} \tilde{x}_k)^2}} \right|, \quad (4)$$

where $\tilde{x}_k$ is the complex sample $x_k$ with the tissue clutter removed by the wall filtering procedure.

In this experiment we calculate distributions of correlation for signals with known types to assess the reliability of the introduced rule and construct maps of correlation for phantoms and *in vivo* data to recognize the presence and type of Twinkling in the volunteers.

RESULTS AND DISCUSSION

**Experiment I**

Twinkling was observed on different objects: on kidney stones submerged in water, agar gel and ethanol, on rusty metal, wooden rods, polystyrene pin targets, on grayscale targets in commercially available phantoms. We tried using different ultrasound probes and machine settings like did the authors of the papers described in section 'Dependence on machine parameters'. We found that Twinkling strongly depends on a tissue-mimicking material or a phantom filler (see Figure 1): in rubber-based material Twinkling appeared only at low PRF (usually <1kHz) in CFI and power Doppler gradually declining with increasing frequency, while in water-based fillers there was no considerable dependence on PRF; the spectra from Twinkling areas for these materials also looked differently: we registered a broadband one (Figure 2a) in water-based materials especially on objects with uneven surfaces; a narrowband spectrum (Figure 2b) was more likely to appear on grayscale targets in rubber-based phantoms as well as on objects submerged in ethanol. The narrowband spectrum was primarily seen at low PRF and depended on wall filter type, while the broadband one seemed to be indifferent to RPF and almost indifferent to the wall filter.

Weinstein et al. (2002) suggested using an additional vibration source to generate Twinkling. We applied the additional vibration source with a variable frequency and observed the narrowband spectrum. The Doppler spectrum seemed to change in direct proportion to the alterations of the source frequency (Figure 2c).

The kidney stone of Volunteer 1 exhibited the broadband spectrum (Figure 2d) along with some tissue motion artifacts.

In this experiment, through observation of pictures obtained with standard ultrasound machines in common Doppler modes we identified two types of characteristic behavior of Twinkling Artifact: it is PRF-indifferent and has the broadband spectrum or appears only at low PRF and has the narrowband spec-trum. In our further discussion, we will refer to it as the 1st and 2nd type of Twinkling, respectively.

**Experiment II**

In this experiment the theoretical signals are compared to the empirical ones. The theoretical signals derive from the mathematical model (2) and presented in Figure 3a-j; the empirical signals were observed in phantoms and given in Figure 3k-t. All the received signals can be divided into four categories:

– <u>signals from stationary targets</u> can be described as $x_k = P_k e^{i\varphi_0} B_k$ where both $P_k$ and $B_k$ are constant for the given location and beam parameters. The exemplary representation of such a signal in complex plane is dot (Figure 3a) and two lines in Cartesian plane in Figure 3b; its counterpart obtained from a non-moving target in a phantom is given in Figure 3k and l and contains slight deviations from a linear form caused by ever-present noise;

– <u>signals from flowing targets</u> like erythrocytes would contain a motion induced Doppler shift $\omega_A$ and, therefore, described as $x_k = P_k e^{i\varphi_0} A_k e^{i\omega_A t_k}$. If dots were described by $A e^{i\omega_A t_k}$ alone, they would be located on the arc of a circle with radius $A$ and center at zero with the arc length proportional to the Doppler shift $\omega_A$ and for sufficiently larger Doppler shifts there would be several rotations around the center of the complex coordinate system and harmonic oscillations in the Cartesian plane. Since targets move in respect to the spatial position of the scanning beam, the instantaneous envelop $P_k$ also changes causing trajectory of the dots to stray as seen in Figure 3c and d. We observed many signals of this type in the vessel area of Gammex 1430 LE Mini-Doppler Flow System; an example is given in Figure 3m and n;

– <u>chaotic signals</u> are those whose amplitude changes from pulse to pulse without an observable pattern; the most known physical source of these is cavitation producing broadband noise-like reflections given in Figure 3e, f, o and p from the theory and physical observations;

– <u>signals from oscillating targets</u>, an example of which according to the mathematical model is given in Figure 3g-j, have a non-constant Doppler shift: it increases, decreases, changes sign; therefore in the complex (Figure 3q) plane the dots are on the arc which is usually quite short and after the filtering procedure this arc jumps to the center of the complex coordinate system (Figure 3s) where it can be closely approximated with a straight line. Signals, as depicted in Figure 3q-t, were observed in the known areas of the $2^{nd}$ type of Twinkling.

In this experiment, we visually analyzed thousands of signals, all falling in the described categories. Agreement between empirical signals observed in phantoms and signals predicted by theoretical model proofs adequacy of the latter.

**Experiment III**

A typical signal corresponding to the $1^{st}$ type of Twinkling is presented in Figure 3o and p: it is a strong echo with randomly distributed phase and amplitude. As for the $2^{nd}$ type of Twinkling signals, according to the mathematical model (2) and Figure 3g-j and q-t, the real and imaginary parts have either identical initial phase shift or their initial phases are off by 180 degrees. In the flow mapping, Doppler power is often used as a threshold to separate blood echoes from the noise. $1^{st}$ and $2^{nd}$ type of Twinkling signals easily exceed the power threshold and are being mapped along with the flow resulting in the appearance of Twinkling Artifact.

Given the distinct appearance of the diagrams of signals, relating to both types of Twinkling, a correlation-based criterion can be introduced and used for their differentiation from each other. Figure 4 presents a distribution of the correlation (4) of signals obtained with the research ultrasound machine from the areas with known types of Twinkling and form the flow area. Altogether we analyzed 4 million signals for the purpose of calculating this distribution to make it statistically valuable. The signals of the 2$^{nd}$ type of Twinkling are on the right as they have a considerable correlation between real and imaginary parts. Signals with a high absolute value of cross-correlation between real and imaginary parts could be observed if the imaged object would vibrate relative to surrounding tissues. For such vibrations to occur we need some energy from either an external acoustic source (Weinstein et al. 2002), a pulse itself (Kargel et al. 2003; Liu et al. 2013), or an internal physiological source, e.g., a heartbeat and breathing. The 1$^{st}$ type of Twinkling typically has a low correlation due to its erratic nature. Flow signals also have a low level of correlation between real and imaginary parts.

As it is evident from Figure 4, the correlation (4) enables us to differentiate the 1$^{st}$ and 2$^{nd}$ types of Twinkling; however, as it is shown in (Leonov et al. 2018c), this differentiation is of a probabilistic nature. Wide areas under the curves can be explained by the presence of an additive noise, which always exists in superposition with the other signal components and cannot be eliminated from them.

Lu et al. (2013) gave strong evidence in favor of crevice microbubbles and cavitation as the cause of Twinkling. Cavitation microbubbles act as random scatterers with their continual vibrations and explosions creating a random Doppler shift between pulses. There were other hypotheses trying to explain random phase fluctuations, namely, Rahmouni et al. (1996), who were further complimented by Kamaya et al. (2003). Yet they both failed to provide a complete and coherent explanation since Rahmouni did not clearly state the cause of Doppler frequency shift and Kamaya justified this shift with a phase jitter, which by an unknown reason should be magnified when ultrasound field interacts with a rough surface.

Lu et al. (2013) proposed microcavitation as the main reason for Twinkling in human subjects. If it is so indeed then correlation in the area of Twinkling should be low. To validate this assumption, three sets of *in vivo* data were analyzed. All sets were captured in a raw radiofrequency form.

Figures 5-7 contain the maps of power and correlation superimposed on grayscale frames with the following color palettes: green to yellow is for the absolute value of correlation between real and imaginary parts of signals; brown to purple represents the power of the Doppler signal. We picked the mentioned colors in order to avoid confusion with the traditional CFI.

The first set contains data from the neck area of Volunteer 1 with a couple of wide vessels and calculus without a strong acoustic shadow in the B-mode image (Figure 5a) making it hard to notice in presence of strong reflecting tissues. There is a correlation map in Figure 5b showing relatively strong correlation only on tendons at the top of the frame and weak levels of correlation in the area of calculus. On the other hand, the power map of the Doppler signals in Figure 5c shows very large power on the calculus. On the power map blood vessels should also be seen, however in this case the Doppler power induced by blood flow is much weaker than the power of Twinkling, signals with relatively weak power are not seen on the map. These findings absolutely support Lu's assumption stating that Twinkling originated from cavitation: cavitation signals due to their erratic nature should have strong power and weak correlation. Figure 5d contains a reference image acquired with Sonoace 8000 EX Prime: both the vessels and the Twinkling are present.

The second set of data is from the kidney of Volunteer 2. This time in Figure 6b correlation on the stone is sufficiently strong. The power in Figure 6c is almost ten times weaker than in Figure 5c. The decrease in power can be attributed to a greater depth; however, the high level of correlation on the stone clearly indicates its vibrations. In this case, microcavitation is not the primary cause of Twinkling. In spectral Doppler one can see the broadband Twinkling spectrum along with the signs of tissue movements (Figure 2d). Figure 6d pre-sents the frame from Sonoace 8000 EX Prime with the Twinkling Artifact on the stone surface.

The source of the third data set was the gallbladder of Volunteer 3. Again, a substantial level of correlation in Figure 7b indicates that in this case microcavitation is not the dominant cause of Twinkling: apparently, both types of Twinkling are present.

## CONCLUSIONS

This study shows Twinkling Artifact as the result of elastic vibrations and cavitation microbubbles. Having two causes Twinkling Artifact appears differently depending on the prevalence of one cause over the other and therefore can be divided into two types with each having its unique properties making it possible to separate the $1^{st}$ type of Twinkling from the $2^{nd}$ one. Elastic vibrations, as well as cavitation bubbles, can be detected even if they are several microns in size and cannot be seen with an optical microscope. Elastic vibrations depend on such parameters of the ultrasound machine as PRF, wall filter characteristics, power, color priority and can often can be suppressed by elevating the wall filter threshold in conjunction with pulse repetition frequency. Microcavitation mostly depends on the roughness of the surface of stone and the power of ultrasound emission and is highly unlikely to occur in rubber-based tissue-mimicking materials, which are generally used in ATS, Blue Phantom and Kyoto Kagaku ultrasound phantoms. Gammex phantoms contain tissue-mimicking watery jelly, in our experiments, the $2^{nd}$ type of Twinkling was observed on thin nylon strings serving as targets for spatial resolution checkups.

The next step should be to design and introduce a novel diagnostic mode for mineral detection which creation is based on the principles discovered in this study. The conducted analysis of Twinkling Artifact signals has demonstrated that kidney stones, urinary calculi and other dense objects could be detected with higher reliability by implementing specific algorithms. The goal for the future is to improve these algorithms. With our technology a better version of CFI can be created, mapping only blood signals and suppressing redundant Twinkling Signs, or, on the contrary, emphasizing Twinkling Signs and concealing blood signals.

Currently, the pilot versions of these modes exist as program modules for real-time processing inside Sonomed-500 ultrasound machine and non-real-time modules with additional features for analysis and experiments (such as, e.g., a broader range of wall-filtering algorithms, possibility to calculate parameter distributions, etc.) available for personal computers. The mode for stone detection overlays a map on top of a grayscale image, similarly to traditional CFI, the only difference is that it depicts not blood flow, but stones and calculi.

## ACKNOWLEDGEMENTS

The authors wish to thank Tatiana Yakovleva and Lubov Pismeniuk for proof reading the article. This study was funded by the Russian Foundation for Basic Research, project no. 17-01-00601.

FIGURES

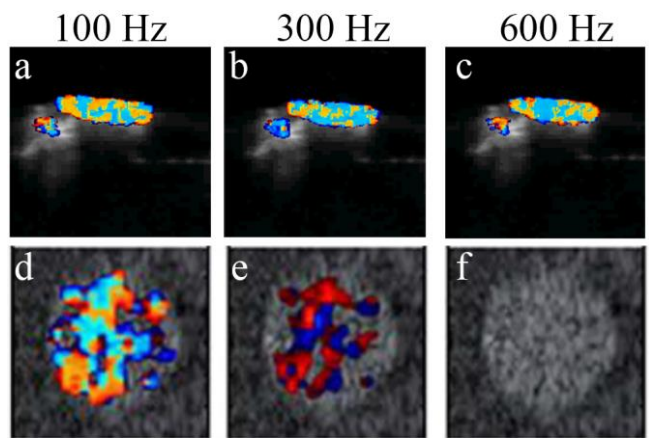

Fig. 1. Twinkling Artifact as a rapid change of shades and colors is seen on an extracted from a kidney stone (a-c) and a circular contrast scattering target inside a phantom (d-f) (frames d-f are borrowed from (Yang et al. 2015)). Images obtained with 100 Hz, 300 Hz and 600 Hz PRF.

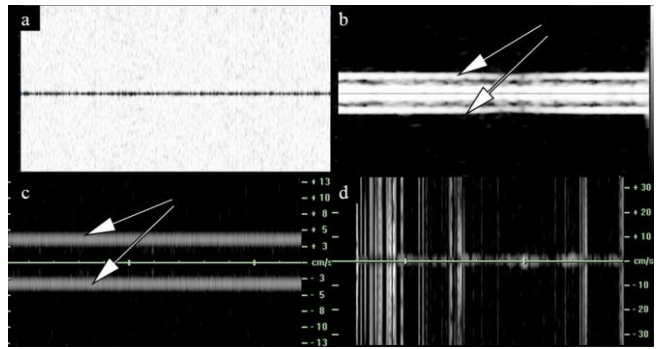

Fig. 2. Images acquired in ultrasound Doppler spectral mode. (a) Typical broad-band spectrum obtained with a custom-designed phantom representing the 1$^{st}$ type of Twinkling at 3 kHz PRF. (b) Narrowband spectrum obtained at 500 Hz PRF with a commercial breast elastography phantom representing the 2$^{nd}$ type of Twinkling. Arrows point at the spectral harmonics associated with the frequency of vibrations. (c) The 2$^{nd}$ type of Twinkling registered with the additional 250 Hz vibration source at 2 kHz PRF. (d) *In vivo* spectrum observed in kidney stones at 5 kHz PRF containing the 1$^{st}$ type of Twinkling and unfiltered tissue signals.

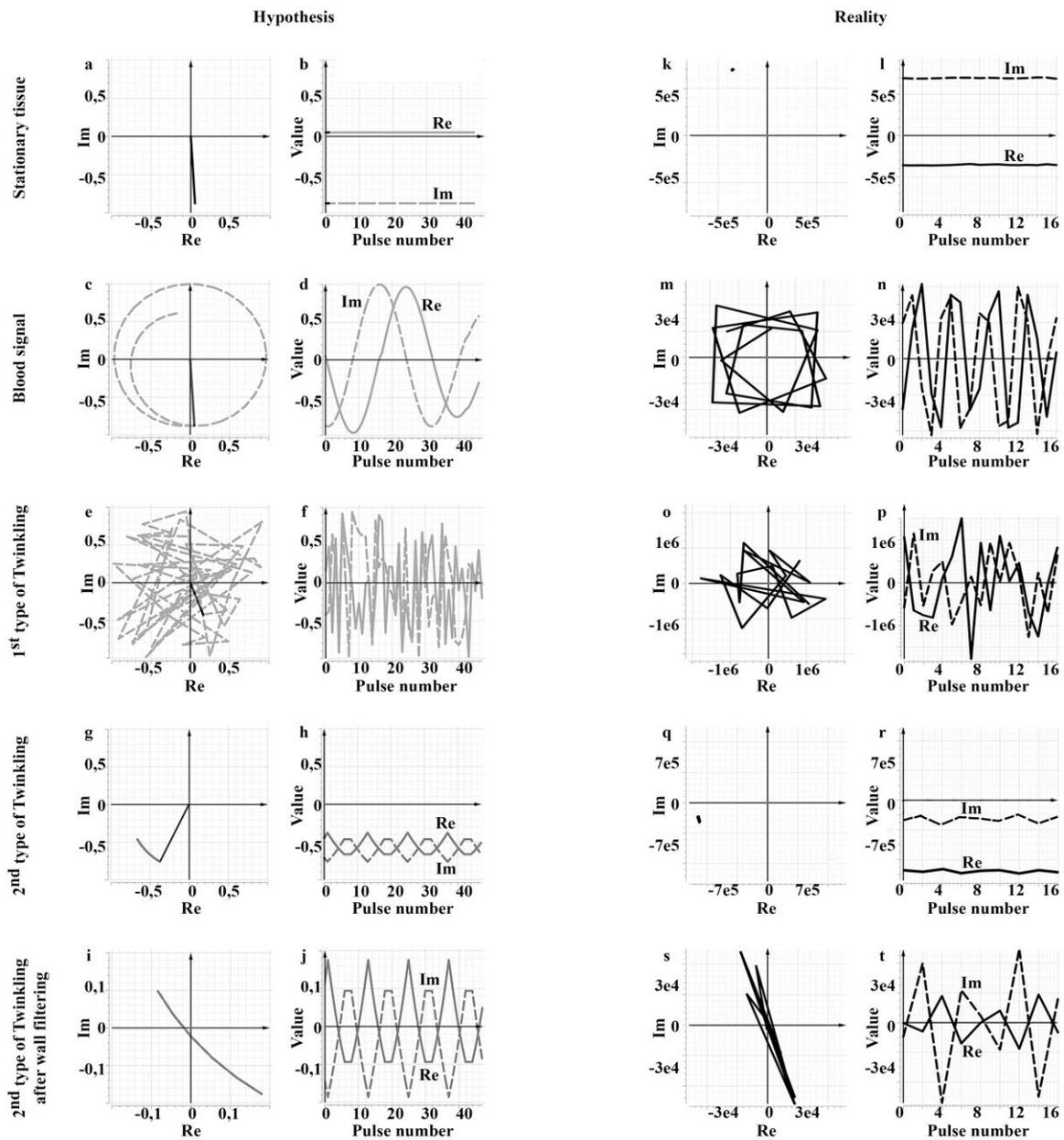

Fig. 3. Typical Doppler signals: signals (a-j) were predicted by the equation (2); signals (k-t) were observed in phantoms. Arbitrary units are used along the axes 'Re', 'Im' and 'Value'; along the 'Pulse number' axis is the consecutive number of the current pulse within the Doppler sequence. The real part refers to the original signal; the imaginary part refers to its Hilbert transform.

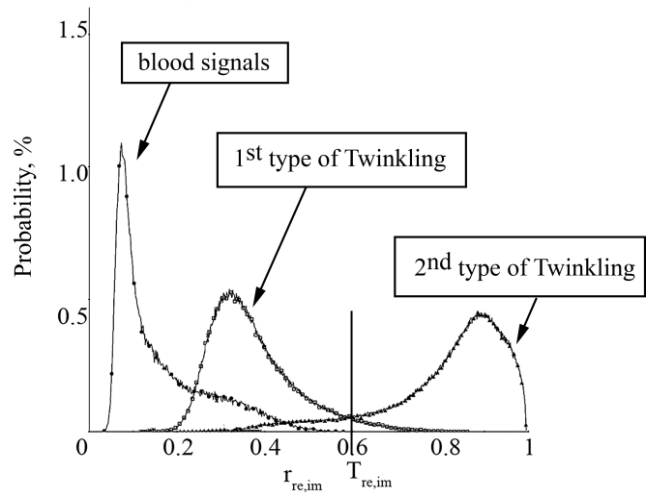

Fig. 4. Distribution of absolute values of correlation between the real and imaginary parts of the typical blood signals and the signals from the regions of both types of Twinkling.

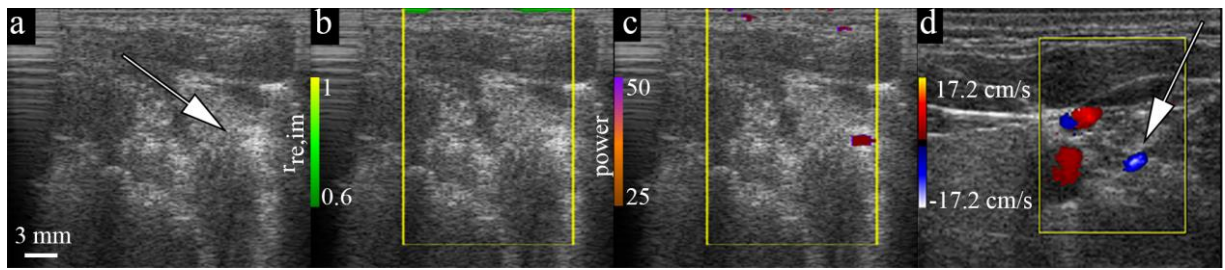

Fig. 5. Picture (a) corresponds to the grayscale image of the neck area; (b, c) – to the same frame with the superimposed real-imaginary correlation map and power map, respectively. Maps demonstrate relatively low correlation levels and strong power typical for the predominance of cavitation in Doppler sequences. Image (d) contains the CFI frame from the reference ultrasound machine demonstrating the presence of Twinkling Artifact. Arrows point at the calculus.

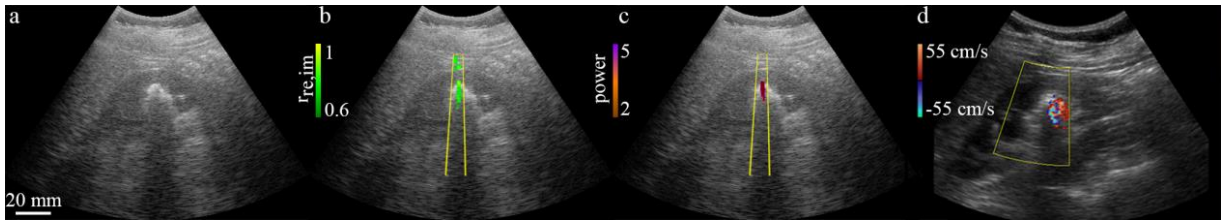

Fig. 6. Experiments with *in vivo* kidney data. Sonogram (a) reveals the kidney stone producing acoustic shadowing. In image (b), the correlation levels are relatively high; in image (c), the power is relatively weak demonstrating the predominance of elastic vibrations in Doppler sequences. Image (d) is the reference CFI frame demonstrating the presence of Twinkling on the kidney stone.

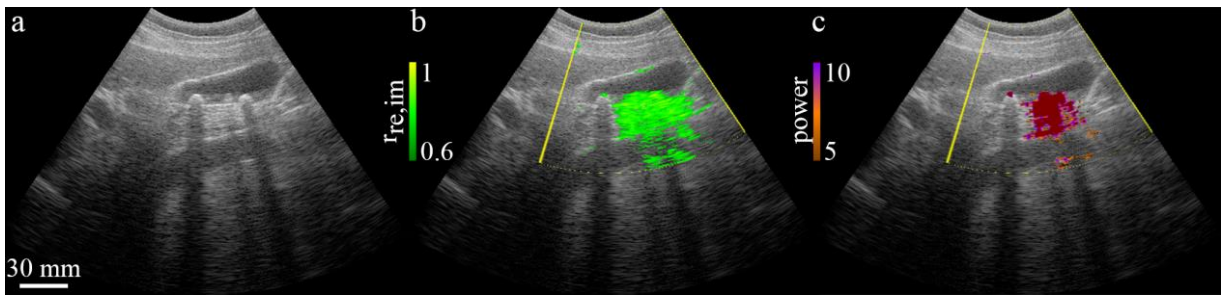

Fig. 7. Experiments with *in vivo* gallbladder data. The gallbladder contains stones (a). Strong correlation levels in image (b) and moderate power in image (c) indicate the predominance of the $2^{nd}$ Type of Twinkling.

TABLE

Table 1. Characterization of dependence on machine parameters by different authors: ↓ – Twinkling decreases when parameter increases; ↑ – Twinkling increases when parameter increases; 0 – Twinkling was indifferent to changes of parameter values; Δ – uncertain; ⊙ – focus on target

| Study | CFI gain | Color priority | Carrier frequency | B gain | Focal position | PRF | Wall filter |
|---|---|---|---|---|---|---|---|
| Behnam et al. 2010 | | | | | ⊙ | | |
| Choi et al. 2014 | ↑ | ↑ | ↓ | ↓ | ↓ | ↓ | ↓ |
| Clarke et al. 2016 | | | | | ↓ | 0 | |
| Gao et al. 2012 | | | ↓ | | | | |
| Haluskiewicz et al. 2017 | ↑ | ↑ | | ↓ | ↓ | | |
| Kamaya et al. 2003 | | Δ | | Δ | | | |
| Lee et al. 2001 | | | | | ↓ | 0 | |
| Louvet 2006 | ↑ | | | | | | |
| Lu et al. 2013 | | | ↓ | | | | |
| Naito et al. 2014 | | | | | | ↓ | |
| Rahmouni et al. 1996 | | | 0 | | 0 | 0 | 0 |
| Shabana et al. 2009 | ↑ | ↑ | 0 | | 0 | 0 | 0 |
| Wang et al. 2011 | ↑ | ↑ | | | ↓ | Δ | ↑ |
| Weinstein et al. 2002 | | | | | | ↓ | |
| Yang et al. 2015 | | | | | | ↓ | |